\documentclass[useAMS,usenatbib]{mn2e}
\usepackage{aas_macros,graphicx,times,multirow}
\usepackage[]{color}
\title[Time-variable emission line in \src]
  {A time-variable, phase-dependent emission line in the X-ray spectrum of the isolated neutron star \src}
\author[De Luca et al.]
{A.~De Luca$^{1,2,3}$ \thanks{E-mail: deluca@iasf-milano.inaf.it},
D.~Salvetti$^{4,1,3}$, A.~Sartori$^{4,1,3}$, P.~Esposito$^{5}$, A.~Tiengo$^{2,1}$,
\newauthor
S.~Zane$^{6}$, R.~Turolla$^{7,6}$, F.~Pizzolato$^{1}$, R.~P.~Mignani$^{6,8}$,  P.~A.~Caraveo$^{1,3}$,
\newauthor 
S.~Mereghetti$^{1}$, and G.~F.~Bignami$^{2,1}$
\smallskip\\
$^1$INAF -- Istituto di Astrofisica Spaziale e Fisica Cosmica - Milano, via E.~Bassini 15, I-20133 Milano, Italy\\
$^2$Istituto Universitario di Studi Superiori, Viale Lungo Ticino Sforza 56, I-27100 Pavia, Italy\\
$^3$Istituto Nazionale di Fisica Nucleare, Sezione di Pavia, Via Bassi 6, I-27100 Pavia, Italy\\
$^4$Universit\`a degli Studi di Pavia, Dipartimento di Fisica Nucleare e Teorica, Via Bassi 6, I-27100 Pavia, Italy\\
$^5$INAF -- Osservatorio Astronomico di Cagliari, localit\`a Poggio dei Pini, strada 54, I-09012 Capoterra, Italy\\
$^6$Mullard Space Science Laboratory, University College London, Holmbury St. Mary, Dorking, Surrey RH5 6NT, UK\\
$^7$Dipartimento di Fisica e Astronomia, Universit\`a degli Studi di Padova, Via Marzolo 8,
I-35131 Padova, Italy\\
$^8$Kepler Institute of Astronomy, University of Zielona G\'ora, Lubuska 2, 65-265, Zielona G\'ora, Poland
}
\date{Accepted \ldots. Received \ldots; in original form \ldots}
\pagerange{\pageref{firstpage}--\pageref{lastpage}} \pubyear{2011}

\def\LaTeX{L\kern-.36em\raise.3ex\hbox{a}\kern-.15em
    T\kern-.1667em\lower.7ex\hbox{E}\kern-.125emX}

\def\xmm {\emph{XMM-Newton}}
\def\cxo {\emph{Chandra}}

\def\src {RX\,J0822--4300}

\def\nh {$N_{\rm H}$}

\begin{document}

\label{firstpage}
\maketitle
\begin{abstract}
\src\ is the Central Compact Object associated with the Puppis A supernova remnant.
Previous X-ray observations suggested \src\ to be a young neutron star with a weak dipole
field
and a peculiar surface temperature distribution dominated by two antipodal spots with different
temperatures and sizes. An emission line at 0.8 keV was also detected. We performed a
very deep (130 ks) observation  with \xmm, which allowed us to study in detail the phase-resolved
properties of \src. Our new data
confirm the existence of a narrow spectral feature, best modelled as an
emission line,
only seen in the `Soft' phase
interval -- when the cooler region is best aligned to the line of sight.
Surprisingly, comparison of our recent observations to the older ones
yields evidence for a variation in the emission line component, which
can be modelled as a decrease in the central energy from $\sim$$0.80$ keV in 2001
to $\sim$$0.73$ keV in 2009--2010. The line could be generated via
cyclotron scattering of thermal photons in an optically thin layer
of gas, or -- alternatively -- it could originate in low-rate accretion
by a debris disk. In any case, a variation in energy, pointing to
a variation of the magnetic field in the line emitting region,
cannot be easily accounted for.
\end{abstract}

\section{Introduction}
Central Compact Objects (CCOs) in Supernova Remnants (SNRs) are a handful of
about 10 point-like, thermally-emitting X-ray sources located close to the
geometrical centres of non-plerionic supernova remnants, with no
counterparts at any other wavelength. CCOs are supposed to be
young, isolated, radio-quiet neutron stars
\citep[see][for a review]{deluca08}.

While the first discovered CCO (the one in the RCW103
SNR) turned out to be a unique object \citep{deluca06},
results of X-ray
timing on a sub-sample of sources with fast periodicities
have recently set very useful constraints for a general picture
of CCOs as a class. Analysis of
multi-epoch \xmm\ and \cxo\ observations of
1E 1207.4--5209 inside G296.5+10.0 and CXOU
J185238.6+004020 at the centre of Kes 79 ($P=424$ ms and $P=105$ ms,
respectively) yielded unambiguous evidence for very
small period derivatives
\citep{halpern10,halpern11}.
This implies, under standard rotating dipole assumptions, characteristic ages
exceeding the age of the host supernova remnants by more than 3 orders
of magnitude, as well as very small ($\la$$10^{11}$ G) dipole magnetic fields.
This points to an interpretation of such sources as `anti-magnetars' -- weakly magnetised isolated
neutron stars, born with a spin period almost identical to the currently observed one.

Such a picture has been recently strengthened by the discovery of 112 ms pulsations from \src, the
CCO in the Puppis A SNR \citep{gotthelf09}, in two archival \xmm\ datasets collected in 2001,
previously used by \citet{hui06} for a comprehensive spectral analysis of the neutron star and of
the surrounding SNR. \citet{gotthelf10}, using a 2010 \cxo\ observation,
set a $2\sigma$ upper limit of $3.5\times10^{-16}$ s s$^{-1}$ to the period derivative,
corresponding to a dipole magnetic field $B<2\times10^{11}$ G, and to a characteristic age
$\tau_\mathrm{c}>5$ Myr, much larger than the age of the host SNR \citep[3.7 kyr][]{winkler88}.
Thus, \src\ can be included in the anti-magnetar family.

The X-ray emission properties of \src\ are very peculiar. \citet{gotthelf09} report on a
$180^{\circ}$ shift in the phase of the pulse peaks occurring abruptly at $\sim$$1.2$ keV. This has
been interpreted as due to the existence of two antipodal hot spots on the star surface, with two
different temperatures (`warm' and `hot') -- indeed seen in the emission spectrum as two different
blackbody components. The star rotation bringing into view or hiding such regions makes the
observed spectrum to change from a soft phase (maximum alignment of the warm spot with the line of
sight) to a hard phase (maximum alignment of the hot spot). Even more intriguing, \cite{gotthelf09}
report on the evidence for an emission line at $\sim$$0.8$ keV, possibly associated with the warm
spot.

Here we report on a very deep \xmm\ observation of \src, performed in 2009 and 2010,
which  allows us to study in detail the phase-resolved emission properties of the neutron star and
to check for any time variability.

\section{observations and data reduction}
Our study is based on a  very deep ($\sim$$130$ ks) observation with \xmm, originally scheduled to
fit into a single satellite orbit (revolution n.1836), starting on 2009 December 17. However,
launch of the {\em Helios 2B} spacecraft required support from \xmm\ ground stations, which
resulted in a $\sim$$50$ ks data gap in the middle of the orbit. The observation was completed
on 2010 April 5. We also included in our study the two archival,
shorter datasets, obtained on 2001 April 15 ($\sim$$29$ ks) and 2001 November 8 ($\sim$$24$ ks),
used by  \citet{gotthelf09} in their previous investigations. A summary of the observations is
reported in Table \ref{obs-log}.

We focus on data obtained by the pn detector \citep{strueder01} of the European Photon Imaging
Camera instrument (EPIC). The detector was operated in the small-window mode
(time
resolution of 5.7 ms, field of view of $4^{\prime}.3 \times 4^{\prime}.3$). The thin optical
filter was used in all observations. EPIC/pn Observation Data Files were processed with the most
recent release of the \xmm\ Science Analysis Software (\textsc{sas} v11) using standard
pipelines.

\section{Timing analysis}\label{timing}

For our timing analysis, we selected the pn source events from a circular region of 30-arcsec
radius,
including only 1- and 2-pixel events (\textsc{pattern} 0 to 4) with the default flag mask. Photon
arrival times were converted to the Solar system barycentre \textsc{tbd} time using the \cxo\
position \citep{gotthelf09}.

Before the analysis presented in \citet{gotthelf09}, the pulsations in \src\
eluded detection for many years, because of a phase shift of about  half a cycle between the nearly
sinusoidal profiles in the soft ($<$1.2 keV) and hard bands. By computing $Z^2_1$ periodograms
\citep{buccheri83short},
we found that the energy bands that maximise the pulsed signal are 0.46--1.17
keV and
1.25--5.12 keV.
The resulting soft and hard pulse profiles of the individual
\xmm\ observations (Table\,\ref{obs-log}) were cross-correlated, shifted and summed to create two
distinct pulse profile templates. Owing to the higher signal-to-noise of the 1.25--5.12 keV
profiles, we carried out the analysis in the hard band, but we checked that the results are fully
consistent with those obtained in the soft band.
The hard
pulse profiles
from temporal segments of the \xmm\ observations were cross-correlated with the template to determine times of
arrival at each
epoch. By means of a phase-fitting analysis (e.g. \citealt{dallosso03}), we measured the periods given in
Table\,\ref{obs-log}
(we also confirm the measurements by \citealt{gotthelf09} for observations A and B).

\begin{figure}
\resizebox{\hsize}{!}{\includegraphics[angle=-90]{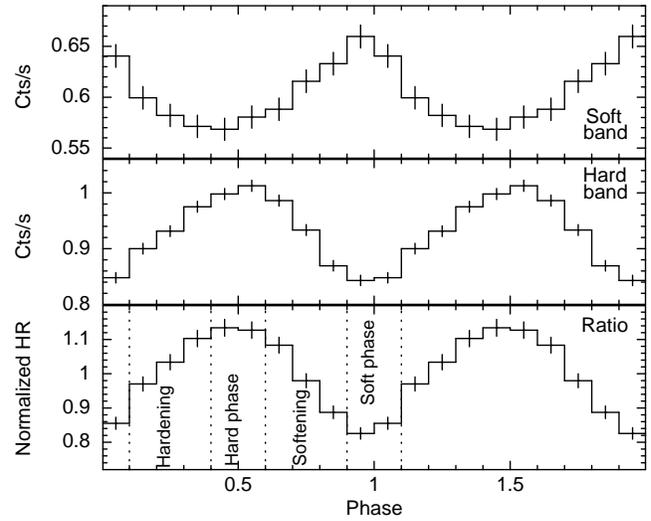}}
\caption{\label{lc} Background-subtracted folded light curves for RX\,J0822--4300 in the
soft energy range (0.46--1.17 keV, upper panel) and in the hard energy range (1.25--5.12 keV, middle panel). The lower panel
shows the hardness ratio (Hard/Soft), normalised to its average value. Phase intervals
used for phase-resolved spectroscopy are also marked.}
\end{figure}

We attempted to obtain a full phase-connected timing solution (i.e. a solution that accounts for all the spin
cycles of the pulsar)
for the longest possible time.
Unfortunately, it was not possible to univocally phase-connect the solution found
for the two adjacent datasets C and D to other observations (the uncertainty
on the phase -- propagated along the large time span to other observations -- largely exceeds 1 cycle).

A linear fit of the periods in Table\,\ref{obs-log} gives for the period derivative $(9\pm12)\times10^{-17}$ s s$^{-1}$,
which translates into 3$\sigma$ limits $-2.7\times10^{-16}$ s s$^{-1} < \dot{P} <4.5\times10^{-16}$ s s$^{-1}$
(in agreement with the limits recently published in \citealt*{gotthelf10}).
The period derivative at the Solar system barycentre results from the \emph{intrinsic} pulsar spin-up/down plus the
contributions due to any external gravitational field and the pulsar proper motion (e.g. \citealt{phinney92}).
In the case of \src, for an assumed distance of 2.2 kpc \citep{reynoso95}, the Galactic contribution is negative and negligible ($\sim$$-2\times10^{-20}$ s s$^{-1}$)
and that produced by the proper motion ($165\pm25$ mas yr$^{-1}$; \citealt{winkler07}) is $1.6^{+0.6}_{-0.4}\times10^{-17}$ s
s$^{-1}$. While the total (Galactic + proper motion) contribution does not impact
significantly on the current limits on the $\dot{P}$ of \src, we note that it is larger than the period derivative measured
for the CCO in Kes\,79 ($\sim$$8.7\times10^{-18}$ s s$^{-1}$;
\citealt{halpern10}).

\begin{table}
\centering
\caption{Journal of the \xmm\ observations. $^{a}$ Mid-point of
  observation. $^{b}$ Time between first and last event. $^{c}$
1$\sigma$ errors in the last digit are quoted in parentheses. $^{d}$
\citet{gotthelf09}.
$^{e}$ This observation was broken into two segments (see text).}
\label{obs-log}
\begin{tabular}{@{}ccccc}
\hline
Dataset & Obs.ID & Date$^{a}$ & Duration$^{b}$& Pulse period$^{c}$ \\
 & & (MJD TBD) & (ks)& (ms)\\
\hline
A & 0113020101 & 52014.471 & 25.1 & \phantom{$^{d}$}112.799\,42(5)$^{d}$\\
B & 0113020301 & 52221.898 & 23.0 & \phantom{$^{d}$}112.799\,44(4)$^{d}$\\
C & \multirow{2}{*}{\phantom{$^{d}$}0606280101$^{e}$} & 55183.064 & 41.9 & \multirow{2}{*}{112.799\,453(2)}\\
D & & 55184.065 & 43.0 & \\
E & 0606280201 & 55291.622 & 42.2 & 112.799\,45(1)\\
\hline
\end{tabular}
\end{table}

\begin{figure}
\resizebox{\hsize}{!}{\includegraphics[angle=-90]{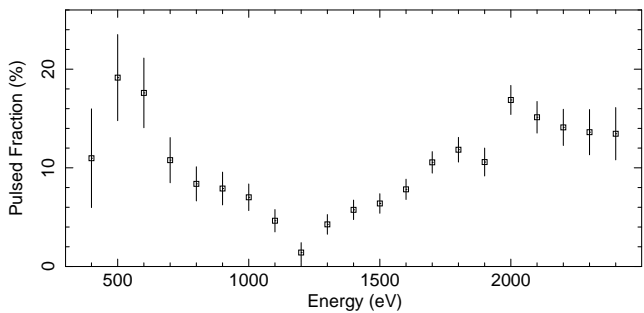}}
\caption{\label{pf} Energy dependence of the pulsed fraction (PF).
The PF was evaluated in overlapping 0.2 keV energy bins, incremented
in 0.1 keV steps, as the ratio between
the number of counts above the minimum and the total number of counts.
Background has been subtracted. 
A clear trend is apparent, with a minimum in the 1.1--1.3 keV energy range.
}
\end{figure}

The phase of the pulse peak is energy dependent (see Figure~\ref{lc}),
with an offset of $0.44\pm0.02$ between the profile as seen at lower ($E<1.17$
keV) and higher energy ($E>1.25$ keV),
the transition occurring quite abruptly at $\sim$$1.2$ keV, consistent
with the findings of \citet{gotthelf09}.
As shown in Figure~\ref{pf}, the pulsed fraction
decreases from $\sim$$15\%$
in the 0.4--0.6 keV energy range to $<$$2\%$ in the 1.1--1.3 keV
range, then grows again to $\sim$$15\%$ at $E>2$ keV, in good agreement
with the model by \citet{gotthelf10}.

\section{Phase-resolved spectroscopy}\label{spec}
Thorough phase-integrated spectroscopy of \src\
has been published by \citet{hui06}. We will focus here on phase-resolved spectroscopy.

\src\ lies in a very complex environment, which makes
background subtraction a critical task. Using phase-integrated data,
we evaluated an optimal selection of source events by maximising
the signal to noise ratio in the 0.3--10 keV  range
as a function of the source extraction radius. The best choice turned
out to be a circle of 17.5$''$ centred at the \textit{Chandra} position.
We extracted the background spectrum from an annular region with radii of 28$''$ and 35$''$ 
centred on the source. Different source and/or
background regions yield consistent results within $\sim$$2\sigma$.

We aligned the photon phases for each observation by cross-correlating
the pulse profiles (see Sect.~\ref{timing}). We generated a combined
event list for first-epoch observations \citep[i.e. data collected in 2001 and used
  by][]{gotthelf09} and one for second-epoch observations (i.e. our new data,
collected in 2009--2010), in view of the large intercurring time span.
We extracted four phase-resolved spectra from both first epoch and second
epoch data, based on  the intervals marked as `Hard phase', `Soft phase', `Hardening'
and `Softening' in Fig.~\ref{lc}.
Phase-resolved spectra were rebinned with at least 100 counts per bin 
and so that the energy resolution was not oversampled by more than a factor of 3.
Response matrices and effective area files for each epoch were generated
by combining (weighted by exposure time) the corresponding files
generated using the \textsc{sas} tools \textsc{rmfgen} and \textsc{arfgen}.
Spectral modelling was performed with the \textsc{xspec} 12.6.0 package in the 0.3$-$4.0 keV energy range.
To account for the effects of
interstellar absorption, we used the \textsc{tbabs} model in \textsc{xspec},
with the abundances by \citet{wilms00}.
We quote errors at 90\% confidence level for a single interesting parameter
unless otherwise specified.

First, we repeated the exercise performed by
\citet{gotthelf09}, fitting a double
blackbody model to our data.
A simultaneous fit to the four phase-resolved spectra
was performed for each epoch. The blackbody normalisations
were allowed to vary as a function of the phase, while the temperatures and
\nh\
were held fixed in order to constrain a single value for all phase ranges.
This model
yields a reduced-$\chi^2$ of of 1.20 for 298 degrees of freedom (d.o.f.) and of 1.23 for 389 d.o.f
for the first and second epoch data, respectively.
Although modulation of
the emitting radii accounts for the bulk of the spectral variation,
structured
residuals in the 0.6--0.9 keV range are apparent
in the Soft phase both in the 2001 dataset \citep[as already reported by][]{gotthelf09}
and in our deeper 2009--2010 dataset, which
confirms the existence of a phase-dependent
spectral feature.

\begin{figure}
\resizebox{\hsize}{!}{\includegraphics[angle=-90]{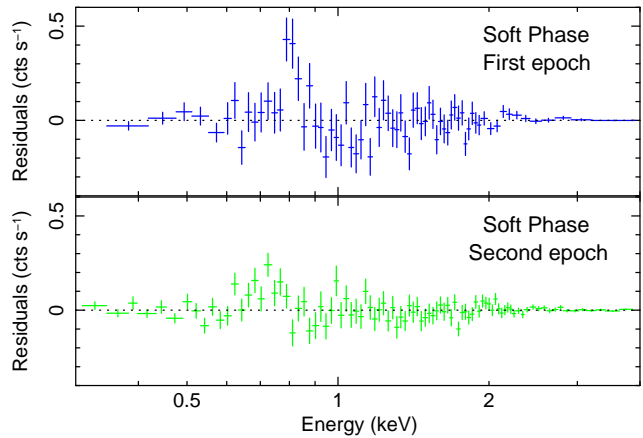}}
\caption{\label{residoldnew}
Residuals (in counts s$^{-1}$) of two-blackbody fits to the Soft phase spectra (see
Fig.~\ref{lc}). A structure in the 0.6--0.9 keV range is seen in both
epochs, although with different shape and intensity. }
\end{figure}

\begin{figure}
\resizebox{\hsize}{!}{\includegraphics[angle=-90]{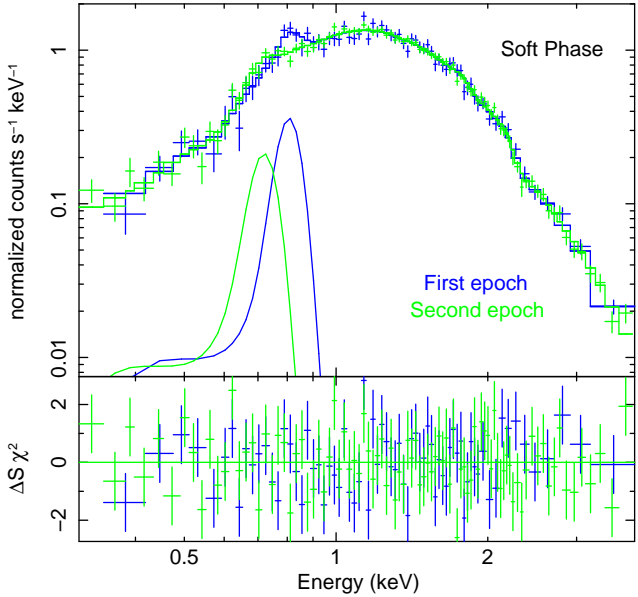}}
\caption{\label{spectraoldnew}
Upper panel: Soft phase spectra for the two epochs.
First epoch: blue. Second epoch: green.
The variation in the 0.6--0.9 keV  range is apparent.
The best-fit model, two blackbody components and a variable
Gaussian emission line, is superimposed.
The line component at both epochs is also shown. The continuum
components do not change as a function of time.
Lower panel: residuals to the best fit
model. The model yields a very good description of the two spectra
($\chi^2_{\nu}=1.02$, 154 d.o.f.).}
\end{figure}

Very interestingly,
while the phase-resolved best fit parameters
do not change as a function
of the epoch -- they can be linked in
a simultaneous two-epoch fit
(more details below) --
the deviation from the continuum in the soft phase
has a somewhat different shape in 2001 with respect to
2009--2010 (see Fig.~\ref{residoldnew}), suggesting
a possible time variability of the spectral feature.
Indeed, such variation is fully apparent when plotting together
the two Soft phase spectra  (see Fig.~\ref{spectraoldnew}).
To quantify the significance of the spectral change
in a model-independent way,
we compared the distributions of the source
events' energies observed in the two epochs
using the Kolmogorov-Smirnov test.
The probability of a statistical fluctuation
producing the apparent difference in the
energy range where the feature is seen ($\sim$0.6--0.9 keV)
turned out to be of $3\times10^{-6}$.

As a second step, in order to model the feature,
we focused on the two Soft phase spectra.
Following \citet{gotthelf09}, we added a Gaussian
emission line to the two-blackbody model.
Indeed, this yields a much better fit with
no structured residuals in the 0.6--0.9 keV range.
As expected, the line component varies as a
function of the time,
its central energy being higher
in the first epoch ($\sim$0.80 keV) than
in the second epoch ($\sim$0.73 keV).
The significance of such
line component was studied by calibrating the F-statistics using
simulations of the
null model (the double blackbody) as suggested by
\citet{protassov02}. Each epoch was studied individually.
Running $10^4$ Monte Carlo simulations, we estimated that
the significance of the line is greater than 99.99$\%$
in the second-epoch spectrum, and greater
than 99.97$\%$ in the first epoch.
Performing a simultaneous fit to the two spectra,
the two-blackbody plus
emission line model yields a reduced-$\chi^2$ of 1.17 (155 d.o.f.)
when all parameters (both continuum and line) are linked, while by allowing
the line central energy to vary between first epoch and second epoch,
the fit improves to a reduced-$\chi^2$ of 1.02 (154 d.o.f.);
see Fig.~\ref{spectraoldnew}.
The chance occurrence probability of such an improvement
is $\sim$$3\times10^{-6}$, as evaluated by an F-test.
Such a result is confirmed by Monte Carlo simulations:
assuming the best fit model of second epoch data,
a line centroid as observed in first epoch could be
obtained by chance with a probability as low as
$\sim$$5\times10^{-6}$.
The fit does not improve significantly by allowing
further line parameters to vary. There is no evidence
for any variation in the
continuum parameters between the two epochs.

We also tried to model the two soft phase spectra using variable
{\em absorption} features. Adding a single Gaussian
absorption line (\textsc{gabs} in \textsc{xspec})
at a higher energy than the one required by the emission line model
($\sim$$1.00$ keV and $\sim$$0.89$ keV in the first and second-epoch,
respectively) yields a much worse fit than the emission line model
(reduced-$\chi^2$ of 1.20 for 154 d.o.f.).
Adding {\em two} lines
at $\sim$$0.97$ keV and at $\sim$$0.63$ keV in the first epoch
and at $\sim$$0.88$ keV and at $\sim$$0.55$ keV in the second epoch
still yields a remarkably worse description of the data
(reduced $\chi^2=1.15$ for 150 dof).
As a further test, we
tried low-$B$ field neutron star atmosphere models for the continuum (\textsc{nsa}, \citealt{zavlin96};
\textsc{nsatmos}, \citealt{heinke06}). As for the blackbody case,
two components with different temperatures are needed. Addition of a variable
Gaussian emission line is still favoured ($\chi^2=1.05$ for 154 dof) with respect to a single variable
absorption line,
as well as to two
variable absorption lines ($\chi^2=1.23$ for 150 dof
for the latter model).

Then, the whole analysis was repeated  for the other phase intervals
(Hard, Softening and Hardening).
No significant improvement in the fit was obtained by adding
an emission line to the two-blackbody model
(the same is true using absorption features).
We assessed that, in each phase interval,
the continuum did not change as a function of the epoch and that
there are no systematic variations between first and second-epoch
in the 0.5--1 keV energy range (such results indicate that the long-term
variability of the feature cannot have an instrumental origin).

As a final step, we performed a simultaneous fit to all the  spectra
based on the results described above.
We used the two-blackbody plus emission line model.
The blackbody normalizations are
phase-dependent, but not
epoch-dependent; the line normalisation is phase-dependent
and its central energy is epoch-dependent.
This yields a reduced
$\chi^2=1.13$ for 691 d.o.f. In such model, the  \nh\ is $(5.0\pm0.1)\times10^{21}$
cm$^{-2}$, the warm blackbody has a temperature $kT_{\mathrm{W}}=265\pm15$ eV
and a radius ranging from 2.27 km (Soft phase) to 2.04 km (Hard phase),
the hot blackbody has a temperature kT$_{\mathrm{H}}=455\pm20$ eV
and a radius ranging from 0.53 km (Soft phase) to 0.65 km (Hard phase).
The line component is narrow ($\sigma\,<\,40$ eV).
In 2001, the line energy is $0.80\pm0.01$ keV and the Equivalent Width (EW) ranges from
$\sim$53 eV in the Soft phase to $<$10--15 eV in other intervals (where
the line is not significant).
In 2009--2010, the line energy is $0.73^{+0.01}_{-0.02}$ keV and the EW ranges from
$\sim$$45$ eV (Soft phase) to $<$12--18 eV in other intervals. The unabsorbed
flux of the line between 0.3 and 10 keV is $\sim$$3\%$
of the flux of the continuum in the same energy range.

\section{Discussion}
Our multi-epoch \xmm\   analysis shows a phase-dependent, {\em time variable}
spectral feature, best modelled as an emission line with
a variable central energy, in the X-ray spectrum of
the `anti-magnetar' candidate \src.

To put such a peculiar result in context, we first note that
our observations
confirm the picture of \src\ as a weakly-magnetised neutron star. Indeed, we  improved the
upper limit on $\dot{P}$, bringing the   dipole
component of the magnetic field down to  $<2.3\times10^{11}$ G
at $3\sigma$ level.
Based on a larger statistics, we also confirm
the geometric model by \citet{gotthelf10}, explaining the
phase-resolved thermal emission with two
antipodal spots of different temperature
(compare e.g. our Fig.~\ref{pf} to their Fig.~6). Lack of any
measurable time variation in the continuum properties
suggests that the warm and hot regions
are intrinsic, persistent features in the thermal map of the star.
To explain such a large surface anisotropy for \src\
(and for CCOs in general) we may consider that a large difference
in intensity could exist between the internal and the external
magnetic fields of the neutron star, as
proposed by \citet{turolla11}
to explain the properties of the low-magnetic field Soft Gamma Repeater
SGR 0418+5729 \citep{rea10}.
In our case, an internal (toroidal + poloidal) field of a few $10^{13}$ G
would be large enough to effectively channel the heat flux from the core
\citep{geppert04,geppert06},
but would not induce crustal fractures
with consequent magnetar-like bursting activity.

The most natural interpretation of the variable emission line is that of
a cyclotron feature produced by electrons.
If its central energy is
associated to the fundamental $e^-$ cyclotron frequency,
the magnetic field in the line emitting region
would be 6--$7\times10^{10}(1+z)$ G (where
$z\sim0.25$ is the gravitational redshift).
This is quite compatible with
the upper limit  from the spin-down rate.
The line is very narrow ($\sigma\leq40$ eV).
If it is a cyclotron line then $\Delta E / E = \Delta B / B$ and
the relative variation of $B$ over the emitting region (conservatively) needs to be $\leq 10\%$.
Thus, the line  should be produced in a very compact region.
A variation of the central energy of the feature
would require  a change either in position of the emitting plasma
within a non-variable magnetic field, or in the intensity
of the magnetic field itself. The lack of changes in the
phase-resolved continuum emission
rules out simpler, purely geometric explanations such as
precession of the neutron star.

To explain the generation of an emission line
in the spectrum of an isolated neutron star, the possibility
of cyclotron scattering of surface thermal photons   
by a geometrically thin, optically thin layer of plasma
could be considered. However, under simple assumptions
(emission from the entire star surface; plane-parallel geometry; pure, conservative scattering), 
a scattering layer would produce an absorption line.
One might 
invoke  a spatially-limited scattering medium, possibly some distance
away from the star surface. Photons coming from the part of the surface not covered by the
layer could be scattered along the line-of-sight giving rise to an emission feature. The value of $B$
derived from the line energy is somehow smaller than the upper limit on the surface field, so the line could
indeed
form at some height in the magnetosphere. A confined medium seems also to
be required by the results of phase-resolved spectroscopy which shows that the emission line is seen mostly
when the cooler spot is into view.
Still, such a picture seems rather contrived, since the nature of the layer and
the mechanism keeping the plasma suspended and confined in a compact
blob remains to be understood.


To ease the problem, an energy source unrelated to the surface
thermal emission should be invoked to excite $e^-$ to higher Landau
levels in the line emitting region.
Indeed, \citet{nelson93}
predicted that for neutron stars accreting at a low rate
($L_{\mathrm{accr}}<10^{34}$ erg s$^{-1}$), and endowed with magnetic fields
of $10^{11}$--$10^{13}$ G,   accreting ions may lose energy to
atmospheric electrons via magnetic Coulomb collisions.
Electrons, excited to high Landau
levels, radiatively decay and part of the cyclotron photons
are expected to escape producing an emission line.
According to \citet{nelson93}, at $B<10^{12}$ G,
the fraction of the accretion-powered flux escaping in the line
is expected to be very small ($\ll$$5\%$), the largest part being
reprocessed and emitted in
a thermal continuum. Thus, one should postulate that the bulk
of the X-ray luminosity of \src\ is accretion-powered, at variance 
with observations. It would require
an accretion rate of   $\sim2\times10^{13}$ g s$^{-1}$ implying -- under
standard relations for propeller spin-down \citep{menou99} -- a $\dot{P}$
value more than 10 times larger than our upper limit. Although the
model by \citet{nelson93} does not fit to our case,
low-level accretion of supernova fallback
material (which cannot be ruled out, based on X-ray timing, as well as on the optical
upper limits set by Mignani et al.2009) could play some role
in generating an emission line in a low-$B$ field atmosphere.
A detailed investigation of such possibility is beyond the scope of this paper.

We will not go into further speculations about the
line emitting mechanism.
We stress that the evidence for time evolution of the spectral
feature is model-independent
and represents to date the first evidence for variability
in an ``anti-magnetar'' candidate.
Likely, such `activity' is related
to a variation of the magnetic field of the  star.
A $\sim$$10\%$ decrease in $\sim$$8$ yr seems vastly too steep to be
attributed to the large-scale dipole field. Possibly,
we are witnessing evolution of a localised multipole component,
dominating close to the star surface. This would hint
at the presence of a large internal field, as proposed
to explain the anisotropic thermal map of the star.
Precise X-ray timing, assessing the 
$\dot{P}$ and measuring the star dipole field, will add
a crucial piece of information.
Coupled to further sensitive phase-resolved
spectroscopy to monitor spectral variability,
this could help to solve the puzzles set by
our results on \src\,
which would have important implications for
the understanding of the nature of CCOs and of their
relations with other families of neutron stars.

\vspace{-2mm}
\section*{Acknowledgments}
Based on observations obtained with \emph{XMM-Newton}, an ESA science mission
with instruments and contributions directly funded by ESA Member States and
NASA. We thank F.~Gastaldello for useful advice about statistics and
A.~Possenti for discussions.  This work was
partially supported by the 
ASI-INAF I/009/10/0 agreement. PE acknowledges financial support from the
Autonomous Region of Sardinia through a research grant under the program PO
Sardegna FSE 2007--2013, L.R. 7/2007.


\bsp

\label{lastpage}

\end{document}